\documentclass[prb,showpacs,twocolumn]{revtex4}
\usepackage{graphicx}
\usepackage{dcolumn}
\usepackage{bm}
\usepackage{amssymb}
\begin{document}
\input psfig.tex
\def\be{\begin{equation}}
\def\ee{\end{equation}}
\def\ba{\begin{eqnarray}}
\def\ea{\end{eqnarray}}
\newcommand{\s}{\sigma}
\title{Effective electron-electron interaction in a two dimensional paramagnetic system}

\author{Juana Moreno and D. C. Marinescu}
\affiliation{Department of Physics, Clemson University, Clemson, SC 29634}
\date{\today}
\begin{abstract}

We analyze the effective electron-electron interaction in a two
dimensional polarized paramagnetic system.
The spin degree of freedom, s, is manifestly
present in the expressions of spin dependent local field factors
that describe the short range
exchange (x) and correlation (c) effects. Starting from the exact
asymptotic values of the local field correction functions for large
and small momentum at zero frequency we obtain self-consistent
expressions across the whole spectrum of momenta. Then, the
effective interaction between two electrons with spins 
s and s' is calculated. We
find that the four effective interactions, 
up-up, up-down, down-up and down-down, are different.
We also obtain their qualitative dependence on the electronic density and
polarization and note that these results are independent
of the approximation used for the local field correction functions
at intermediate momenta.

\end{abstract}

\maketitle

\section{\bf Introduction}

To accomplish spin dependent conduction in electronic devices has
become a very intense quest in recent years. \cite{review} 
The II-VI dilute magnetic semiconductors, like CdMnTe and ZnMnSe, 
seem to be the most promising materials for achieving this goal. 
At low doping levels these systems
exhibit an enhanced Zeeman splitting of the electronic levels that
arises from strong Kondo-like exchange between the embedded Mn
ions and the delocalized electronic states. 
Moreover, the II-VI dilute
magnetic semiconductors are n-type conductors with
relative good mobility and large spin coherence times ($\sim 10
ns$). However, an open question remains as to how the
magnetic interaction between the localized spins and the itinerant
carriers reflects on the transport properties of these structures.
The purpose of this work is to analyze some aspects of this
problem.

 Since the Zeeman
splitting of the electron levels was found to be independent of the
local magnetic environment \cite{crooker95}, a simple model of a
II-VI magnetic semiconductor is a spin polarized electron gas in
the presence of a static magnetic field. 
The field lifts the spin degeneracy and induces an equilibrium
polarization, $\zeta = (n_{\uparrow}-
n_{\downarrow})/(n_{\uparrow} + n_{\downarrow})$.  
The strong magnetic interaction between the itinerant carriers and
the localized spins is reflected in the large value of the
effective gyromagnetic factor, $\gamma^*$, up to hundreds of times
the band value. On account of
the large $\gamma^*$, even low magnetic fields are enough to
produce large polarizations, and it is assumed that $\zeta$ can
vary continuously between $-1$ and $1$ as a function of the static
magnetic field. This approximation integrates out the degrees of
freedom of the static spins under the enlarged value of $\gamma^*$
and focuses on the itinerant carriers and the many body
interaction among them. The latter is independent of the source of
the spin polarization, and we expect our results to maintain their
validity also in the case of a self-consistent magnetic field as
source of the spin polarization, a situation which is consistent with an
itinerant ferromagnet. The model can also be extended to describe
the paramagnetic state of the III-V magnetic heterostructures
where a mostly uniform internal magnetic field is created by
having a magnetic ion density much larger than the
itinerant-carrier density. \cite{mune89,ohno92} However, the small
electronic densities in the III-V based compounds make difficult
the blind use of the paramagnetic electron gas model.

The explicit spin dependence of the electron-electron interaction
becomes manifest when the exchange (x) and correlation (c) effects
of the local Coulomb repulsion are included. At finite
values of the polarization, the many-body short range interaction, which
is density dependent, is different for up and down spin electrons.
A realistic picture of the electronic interaction and its
screening is obtained by using spin dependent local field
correction functions, $G^{x,c}_{\sigma}({\bf q},\omega)$, that
describe the exchange and correlation hole around each electron.
The self consistent treatment of exchange and correlation effects
has proved very important in understanding the physics of normal
metals, \cite{mahan} but to our knowledge it has not been fully
analyzed in spin-polarized systems, where the spin dependence of
the local field corrections becomes manifest. In addition, the
relevance of exchange and correlation effects increases as the
dimensionality of the electron gas is lowered. For example, the
importance of local effects is clearly reflected in the dielectric
function of an unpolarized two-dimensional electron gas, which
becomes overscreened in a wide range of momenta and electronic
density. \cite{pede97} However, calculations where these effects
are absent, such as the conventional random phase approximation,
predict a positive dielectric function for the whole parameter
range. Therefore, we believe it is fundamental to include these
local effects in a complete treatment of the quasi-two-dimensional
diluted magnetic semiconductors.

Obtaining the exact frequency and wave vector dependence of the
local field corrections is a very difficult problem which
remains unsolved even in the case of the unpolarized
electron system. Fortunately, the asymptotic values
of the local field factors 
can be obtained exactly in two limiting cases. At zero frequency
and small wavevectors, sum rules are used to connect the static
limits of the response functions to certain thermodynamic
coefficients. \cite{mar02} For large frequency and large
wavevector, an iterative method generates the exact expressions
for the local field functions up to second order. \cite{niklasson}
This approach uses the equation of motion satisfied by the
Wigner distribution function of the particle density.

Numerical estimates of the response functions of the three
dimensional unpolarized electron gas have shown that local field
factors smoothly interpolate between the asymptotic small and
large wave-vector behavior.\cite{Moro95} This feature is expected
to exist also in the case of a spin polarized system, and,
consequently, we use the exact asymptotic values of
$G^{x,c}_{\sigma}({\bf q},\omega)$ for large and small momentum at
zero frequency \cite{mar02,mar97,pol01} as a starting point in
obtaining their approximate expressions across the whole spectrum
of momentum. Then, $G_{\s}({\bf q},\omega)$ are used to calculate the
spin dependent effective electron interaction potentials, ${{\cal
V}_{\sigma \sigma'}}({\bf q}, \omega)$. We compare two
interpolation schemes to conclude that the effective potentials are
qualitatively independent of the approximation used for the local
field correction functions at intermediate momenta.

In section II, we present the self consistent formulation
of the effective interaction that incorporates the spin dependent
local field corrections following the approach of Kukkonen and
Overhauser \cite{kuk79,zhu86} generalized for a spin polarized
electron system.\cite{yi96} This formalism permits the derivation
of the effective spin dependent scattering potentials experienced
by an electron and  of  the response functions. In section III, we
discuss in detail the procedure used to obtain the approximate
expressions for the local field factors. Section IV presents our
conclusions. The appendix is dedicated to a derivation of the
correlation function of a two dimensional electron gas at the
origin.

\section{\bf Effective interaction potentials}

\begin{figure}
\begin{minipage}{\linewidth}
\includegraphics[width=\textwidth]{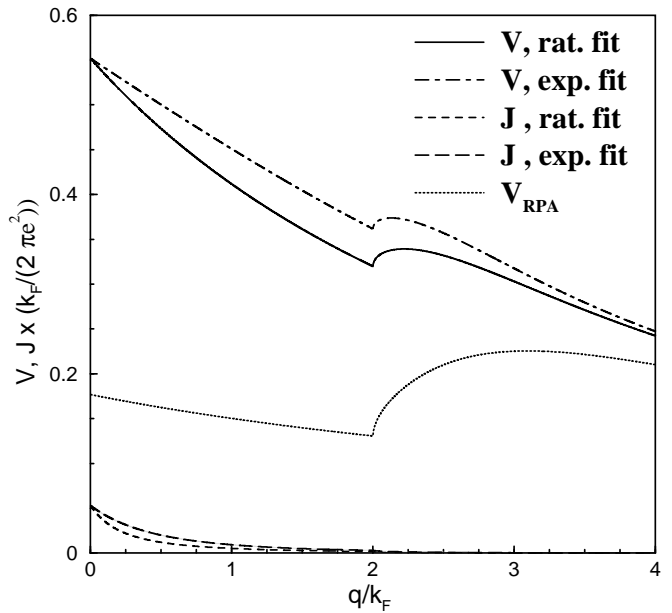}
\end{minipage}
\caption {Normalized static effective interactions
$V({\bf q},\omega=0)$  and $J({\bf q},\omega=0)$ for an unpolarized
electron gas ($\zeta=0$) as  functions of the normalized momentum
${\bf q}/k_F$ for $r_s=4$. Results using a rational (Eq.~(\ref{eq:posrat}))
and an exponential (Eq.~(\ref{eq:posexp}))
fit for the local field correction functions are displayed. Note
the similarity of the results for the two fittings.
Results for the random phase approximation are also displayed.}
\label{fig:2Dpot0}
\end{figure}

\begin{figure}
\begin{minipage}{\linewidth}
\includegraphics[width= \textwidth]{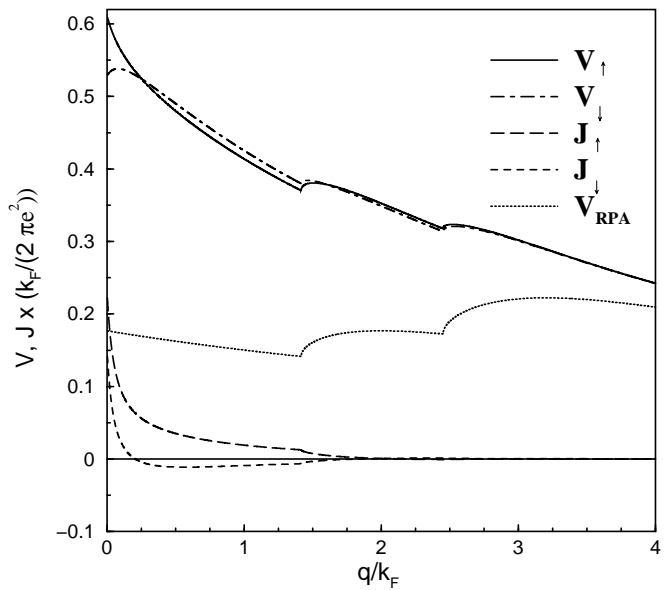}
\end{minipage}
\caption {Normalized static effective interactions
$V_{\sigma}({\bf q},\omega=0)$  and
$J_{\sigma}({\bf q},\omega=0)$ for a polarization of $\zeta=0.5$
as  functions of the normalized momentum ${\bf q}/k_F$ for
$r_s=4$. Results for the random phase approximation are also displayed.
We have used a rational fit (Eq.~(\ref{eq:posrat}))  for the local field
corrections.}
\label{fig:2Dpot5}
\end{figure}

\begin{figure}
\begin{minipage}{\linewidth}
\includegraphics[width=\textwidth]{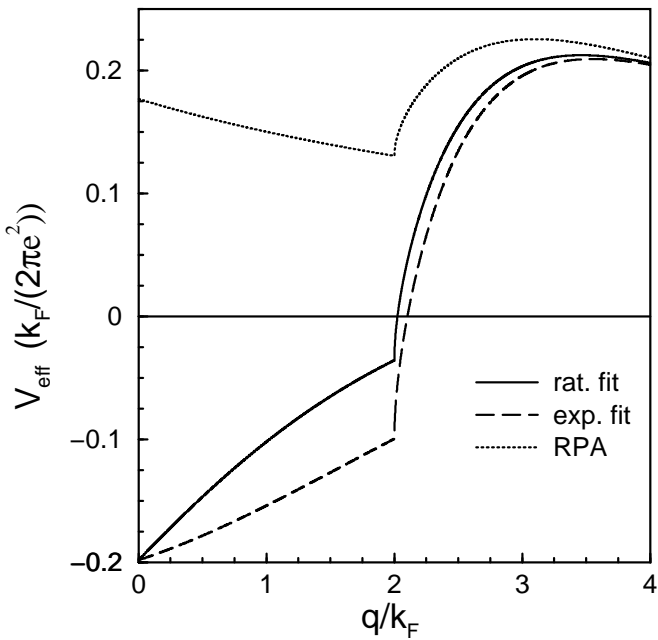}
\end{minipage}
\caption {Normalized effective interaction
$V_{eff}({\bf q})$ for an unpolarized
electron gas ($\zeta=0$) as  a function of the normalized momentum
${\bf q}/k_F$ for $r_s=4$. Results using a rational (Eq.~(\ref{eq:posrat}))
and an exponential (Eq.~(\ref{eq:posexp}))
fit for the local field correction functions as well as
results for the random phase approximation are  displayed.
Note that the RPA interaction is the same as that in Fig.~\ref{fig:2Dpot0}.}
\label{fig:2DVeff0}
\end{figure}

\begin{figure}
\begin{minipage}{\linewidth}
\includegraphics[width= \textwidth]{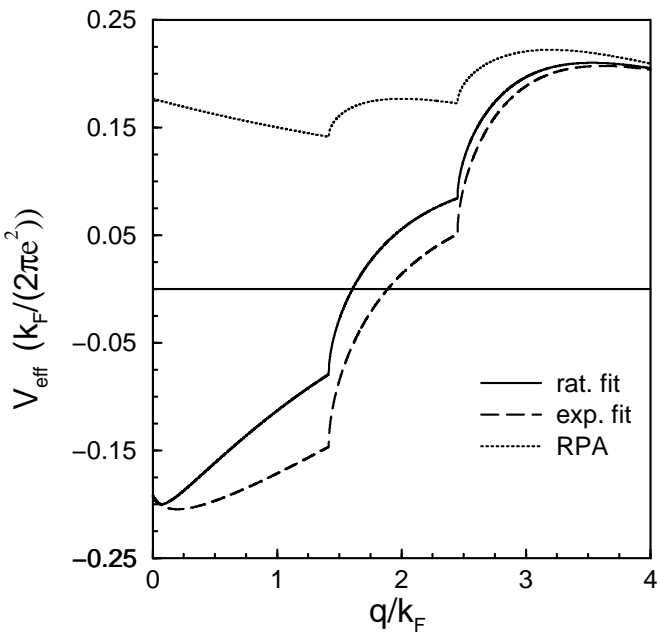}
\end{minipage}
\caption {Normalized effective interaction
$V_{eff}({\bf q})$  for a polarization of $\zeta=0.5$
as a function of the normalized momentum ${\bf q}/k_F$ for
$r_s=4$. Results using a rational (Eq.~(\ref{eq:posrat}))
and an exponential (Eq.~(\ref{eq:posexp}))
fit for the local field correction functions as well as
results for the random phase approximation are  displayed.
The RPA curve is the same as that in Fig.~\ref{fig:2Dpot5}.}
\label{fig:2DVeff5}
\end{figure}

The simplest approximation of the effective interaction in the
electron gas is the random phase approximation (RPA), which
incorporates the screening but ignores the exchange and
correlation effects. Therefore, all spin dependence is lost. A more
realistic description, which considers the short range Coulomb
interaction effects, was proposed by Hubbard, \cite{Hubb57} whose
expression for the dielectric constant introduces the local field
correction, a wavevector dependent function that describes the
difference between  the real particle density and its mean field (RPA)
counterpart.

The microscopic origin of the local field corrections was
elucidated by Kukkonen and Overhauser. \cite{kuk79} In their model
of the electron-electron interaction, the exchange (x) and
correlation (c) are explicitly incorporated by considering
associated local field factors, $G (\bf{q},\omega)$, in the
expression of the effective potential experienced by one electron
in the presence of all the rest. In a self-consistent formulation,
the details of which will be given below in the case of a spin
polarized electron system, this theory leads to the Hubbard
dielectric function. The method is equivalent to considering the
many body interaction in all orders, as it has been shown by using
diagrammatic techniques. \cite{Vig85}

In the case of a spin polarized system, it is appropriate to
introduce spin dependent local field corrections because the many
body effects are density dependent, and they depend on the spin
directly through the exclusion principle.\cite{yi96} In a direct
generalization of Ref. \onlinecite{kuk79}, the effective potential
$\widetilde{\cal V}_{\sigma \sigma'}$ experienced by an electron
with spin $\sigma$ in the presence of an second electron
with spin $\sigma'$ and charge density $\rho({\bf q})$ 
can be written as: \cite{spin-flip}

\ba
\label{eq:effint}
\widetilde{\cal V}_{\sigma \sigma'}({\bf q},
\omega) = v({\bf q}) \Big\{ [1 - G^{+}_{\sigma}({\bf q},\omega)]
(\rho({\bf q})+
\Delta n({\bf q},\omega)) \nonumber \\
- \sigma G^{-}_{\sigma}({\bf q},\omega)
(\sigma'\rho({\bf q})  + \Delta m_z({\bf q},\omega))
\Big\}
\ea
where 
$\Delta n= \Delta n_{\uparrow} +\Delta n_{\downarrow}$ and $\Delta
m_z= \Delta n_{\uparrow} -\Delta n_{\downarrow}$ are the particle
and spin density fluctuations, 
respectively, induced by the presence of the second electron, and
$v(q)$ is the bare Coulomb
interaction, which is equal to $2 \pi e^2/q$ in two dimensions. The local
field functions $G^{\pm}_{\sigma}({\bf q},\omega)$ incorporate the
exchange ($G^x_{\sigma}$) and correlation effects ($G^c_{\sigma
\sigma}, G^c_{\sigma \bar{\sigma}}$), which induce a local decrease
in the electronic density  around an electron of given spin
$\sigma$ compared with its RPA value.~ $G^{+}_{\sigma}({\bf q},\omega)$
is the sum of all same and opposite spin effects, and
$G^{-}_{\sigma}({\bf q},\omega)$ is the difference of same and opposite
spin effects: $G^{\pm}_{\sigma}({\bf q},\omega)=
G^x_{\sigma}+G^c_{\sigma \sigma} \pm G^c_{\sigma \bar{\sigma}}$. 
\cite{same-spin} In brief, Eq.~(\ref{eq:effint})
shows how the electron charge density is decreased from the
corresponding RPA value to account for its own short range
effects.

The role played by $G^{+}_{\sigma}$ and $G^{-}_{\sigma}$ in
the physical properties of the electron gas  is quite different.
For an unpolarized electron gas, $G^{+}$ enters the
electrical response functions, and
$G^{-}$ the magnetic response.
It is well known that in this limit the dielectric
function and the electronic susceptibility can be written down as
functions of $G^+({\bf q})$ and $G^-({\bf q})$ respectively:
\cite{kuk79}
\ba
\label{eq:np.eps}
\epsilon({\bf q},\omega)= 1- \frac{v_q \Pi({\bf q},\omega)}
{1+G^{+}({\bf q}) v_q \Pi({\bf q},\omega)}\\
\label{eq:np.sus}
\chi({\bf q},\omega)= \frac{-\gamma^{*2} \Pi({\bf q},\omega)} 
{1+G^{-}({\bf q}) v_q \Pi({\bf q},\omega)}
\ea
where $\gamma^*$ is the enhanced
effective gyromagnetic ratio and $\Pi=\Pi_{\uparrow
\uparrow}+\Pi_{\downarrow \downarrow}$ is the polarization function.
However, in a polarized gas both local field factors are 
interconnected and appear in
the expressions of the dielectric function and the magnetic susceptibility.

In a linear approximation, the density fluctuations are
proportional to the effective potentials, where the
proportionality coefficients are appropriately defined
polarization functions: $\Delta n_{\uparrow}= \Pi_{ \uparrow
\uparrow} \widetilde{\cal V}_{\uparrow,\sigma'}$ ($\Delta
n_{\downarrow}= \Pi_{ \downarrow \downarrow} \widetilde {\cal
V}_{\downarrow,\sigma'}$). Here $\Pi_{\sigma \sigma'}$ is the
generalized polarization bubble for the non-interacting electron
gas,
\be
\Pi_{\sigma \sigma'}({\bf q},\omega)= \frac{1}{V}
\sum_{\bf k} \frac{n_{\bf k, \sigma} - n_{\bf k +\bf q, \sigma'}}
{\omega + \xi_{\bf k \sigma} - \xi_{\bf k+ \bf q \sigma'}}
\label{eq:polarization}
\ee
where $\xi_{\bf k \sigma}=
\epsilon_{\bf k} + \mbox{sign}(\sigma) \gamma^* B$ is the single
particle energy in the static magnetic field B,
$n_{\bf k, \sigma}$ is
the occupation function, and V is the volume of the system.
Eq.~(\ref{eq:polarization}) generates the
well known expressions for the polarization function of the
non-interacting electron gas when the
usual simplifications are considered: zero temperature and
a parabolic energy dispersion, $\epsilon_{\bf k} = \hbar^2 k^2/2m^*$,
where $m^*$ is the electronic band mass.
Neither approximation is expected to modify our results.

Therefore,  in two dimensions the
polarization operator for the complex frequency $i\omega$ is:
\cite{Stern67}
\ba
\Pi_{\sigma \sigma}(\tilde{q},
i\widetilde{\omega})= \frac{k^2_F}{E_F} \Big( - \frac{1}{4
\pi}\Big)
\Big\{ 1 -\frac{\widetilde{k}_{F\sigma}}{\tilde{q}}  \nonumber \\
\Big[ (i \widetilde{\omega} +\tilde{q}^2) \sqrt{\Big(\frac{1}{2
\tilde{q} \widetilde{k}_{F\sigma}}\Big)^2 -\Big(\frac{1}{i
\widetilde{\omega} +\tilde{q}^2}\Big)^2} + c.c \Big] \Big\}
\ea
where $\tilde{q}=q/k_F$ is the  normalized momentum,
$\widetilde{k}_{F\sigma}=k_{F\sigma}/k_F$
the normalized Fermi momenta of the spin $\sigma$ electronic population,
and $\widetilde{\omega}=\omega/E_F$  the normalized frequency.
The use of normalized variables allows us to
express easily all our results in terms of our only
free parameter: the effective electronic density, or
the ratio between $r_s$
and the effective Bohr radius of the
system, $a^{*}_B=\hbar^2/(m^* e^{*2})$.\cite{numbers}
The retarded polarizability can be obtained by analytical
continuation.

The truly effective interaction potentials,
which are used in the calculation of scattering amplitudes, are obtained
from $\widetilde{\cal V}_{\sigma \sigma'}$ by subtracting the
direct exchange and correlation between the two electrons: the
term $v(q)\rho({\bf q})[ - G^{+}_{\sigma}({\bf q})- \sigma \sigma' 
G^{-}_{\sigma}({\bf q})]$ in
Eq.~\ref{eq:effint}. Thus, the effective interaction can be
expressed as:
\be
{\cal V}_{\sigma,\sigma'}= V_{\sigma'}-
J_{\sigma'} \sigma \cdot\sigma'
\ee
where the first term includes
the bare Coulomb interaction and the interaction mediated by
charge fluctuations, and the second term reflects the interaction
mediated by spin fluctuations. In an unpolarized electron gas, $V$
and $J$ are spin independent. In addition, $V$ depends only on
$G^{+}$ while $J$ depends only on $G^{-}$, as it was shown in Ref.
\onlinecite{kuk79}. However, the imbalance between the two spin
populations in a spin polarized electron gas induces a dependence
on the spin index. Thus, $V_{\sigma}$ and $J_{\sigma}$ are:
\begin{widetext}
\ba
\label{eq:VJeff}
V_{\sigma}&=& [{v_{\bf q}\rho_{\bf q}}/{D({\bf q},\omega)}] \Big\{ 1
+\frac{1}{2}[G^{+}_{\sigma}+G^{+}_{\bar{\sigma}}
+G^{-}_{\sigma}-G^{-}_{\bar{\sigma}}][D({\bf q},\omega)-1 ]
+v_{\bf q} \Pi_{{\bar{\sigma}}{\bar{\sigma}}}
[G^{-}_{\sigma}(1-G^{+}_{\bar{\sigma}})
+G^{-}_{\bar{\sigma}} (1-G^{+}_{\sigma})]\Big\} \\
J_{\sigma}&=&- [{v_{\bf q}\rho_{\bf q}}/{D({\bf q},\omega)}] \Big\{
\frac{1}{2}[G^{+}_{\sigma}-G^{+}_{\bar{\sigma}}
+G^{-}_{\sigma}+G^{-}_{\bar{\sigma}}][D({\bf q},\omega)-1 ]
+v_{\bf q} \Pi_{\bar{\sigma}\bar{\sigma}}
[G^{-}_{\sigma}(1-G^{+}_{\bar{\sigma}}) +G^{-}_{\bar{\sigma}}
(1-G^{+}_{\sigma})]\Big\} 
\ea
where
\hspace{0.1in} $D({\bf q},\omega)= [1-v_{\bf q} (1
-G^{+}_{\uparrow}-G^{-}_{\uparrow}) \Pi_{\uparrow
\uparrow}][1-v_{\bf q} (1 -G^{+}_{\downarrow}-G^{-}_{\downarrow})
\Pi_{\downarrow \downarrow}]-[1
-G^{+}_{\uparrow}+G^{-}_{\uparrow}] [1
-G^{+}_{\downarrow}+G^{-}_{\downarrow}]v^2_{\bf q} \Pi_{\uparrow
\uparrow} \Pi_{\downarrow \downarrow}$.
\end{widetext}

Figs.~\ref{fig:2Dpot0} and~\ref{fig:2Dpot5} 
show the static effective interactions
$V_{\sigma}({\bf q},\omega=0)$ and $J_{\sigma}({\bf q},\omega=0)$
between point-like electrons \cite{Fourier} 
in a two dimensional electron gas as functions
of the normalized momentum $q/k_F$. Fig.~\ref{fig:2Dpot0}  displays the
results for an unpolarized electron gas. Fig.~\ref{fig:2Dpot5} corresponds
to a  $\zeta=0.5$ polarization. We have chosen $r_s=4$, where
$r_s$ is measured in units of the effective Bohr radius of the
system. \cite{numbers} The expressions used for the local field
corrections are discussed in the following section. For
comparison, we also include the RPA effective interaction
potential, $V_{RPA}({\bf q},\omega=0)$, obtained 
from equation~(\ref{eq:VJeff}) by neglecting the local field corrections. Note
that $J_{RPA}({\bf q},\omega=0)=0$, since the interaction mediated
by spin fluctuations is directly proportional to the local
corrections.

Fig.~\ref{fig:2Dpot0} shows clearly the importance of the
local effect in the calculation of physical properties. First,
as we have already mentioned, the spin mediated interaction,
$J_{\sigma}({\bf q},\omega)$, becomes noticeable. Second,
$V_{\sigma}({\bf q},\omega)$ is greatly enhanced at small momenta
in comparison with the RPA prediction. In addition, 
Fig.~\ref{fig:2Dpot5} shows how
both effective interactions split when the electron
gas is  polarized, an effect also absent in the RPA approach.
The splitting of $J_{\sigma}({\bf q},\omega)$ is more pronounced
that the splitting of $V_{\sigma}({\bf q},\omega)$ and, even,
$J_{\downarrow}$ becomes negative for intermediate momenta.

Larger values of the polarization induce larger splitting of the
up and down effective interactions until a spin wave instability
is reached. For the density value we are considering ($r_s=4$),
this instability happens at $\zeta \sim 0.72$. At that value of the
polarization, $J_{\sigma}({\bf q},\omega)$ and $V_{\sigma}({\bf
q},\omega)$ diverge for a certain value of the momenta that
define the periodicity of the spin density wave.

In order to have a complete picture of the effective interactions, we
also discuss the effective interaction which is used to calculate 
the electronic 
selfenergy. It can be derived in a similar manner
using the self-consistent relation established between the density
fluctuations and the effective potential induces by an external charge.
\cite{Mor02} This effective interaction can be written as 
$\displaystyle V_{eff}= \frac{v({\bf q})}{\epsilon({\bf q},\omega=0)}$,
where $\epsilon({\bf q},\omega)$ is the dielectric function of
a polarized electron gas:
\be
\epsilon ({\bf q}, \omega)= 1-
v_{{\bf q}}\frac{\Pi_{\uparrow \uparrow}+
\Pi_{\downarrow \downarrow}+ 2 v_{\bf q} 
(G^{-}_{\uparrow}+G^{-}_{\downarrow})
\Pi_{\uparrow \uparrow} \Pi_{\downarrow \downarrow}}
{\widetilde{D}({\bf q}, \omega)}
\ee
where \hspace{0.1in}
$\widetilde{D}({\bf q}, \omega)=[1+v_{\bf q} 
(G^{+}_{\uparrow}+G^{-}_{\uparrow}) \Pi_{\uparrow \uparrow}]
[1+v_{\bf q} (G^{+}_{\downarrow}+G^{-}_{\downarrow})
\Pi_{\downarrow \downarrow}]-
[G^{+}_{\uparrow}-G^{-}_{\uparrow}] [G^{+}_{\downarrow}-G^{-}_{\downarrow}]
v^2_{\bf q} \Pi_{\uparrow\uparrow} \Pi_{\downarrow \downarrow}$.

Fig.~\ref{fig:2DVeff0} displays $V_{eff}({\bf q})$ for an unpolarized 
electron gas with the same density as before ($r_s=4$). 	
Since  $V_{eff}({\bf q})$ represents the electrostatic potential seen 
by a spinless test charge,\cite{kuk79} it is very illustrative to compare 
it with the RPA result, 
$\displaystyle V_{RPA}= \frac{v({\bf q})}{\epsilon_{RPA}({\bf q},\omega=0)}$.
\cite{RPA} 
We find that by including the local factors the effective interaction becomes
overscreened for ${\bf q}$ less than a critical value, 
${\bf q}_c$, which depends on the electronic density and the polarization.  
Similar results for zero polarization were reported previously. \cite{bul96}
The critical value of the momentum ${\bf q}_c$ depends weakly on
the approximation used for the local field factors.
Therefore, the effective interaction with local field corrections 
is very different from   
the RPA effective interaction which is always positive.
Fig.~\ref{fig:2DVeff5} displays $V_{eff}({\bf q})$ for an
electron gas with $\zeta=0.5$. The overscreened region is 
also present, and the singularities at the Fermi momenta of
the majority and minority spin populations are clearly seen.

\section{\bf Spin Dependent Local Field Corrections}

\begin{figure}
\begin{minipage}{\linewidth}
\includegraphics[width=\textwidth]{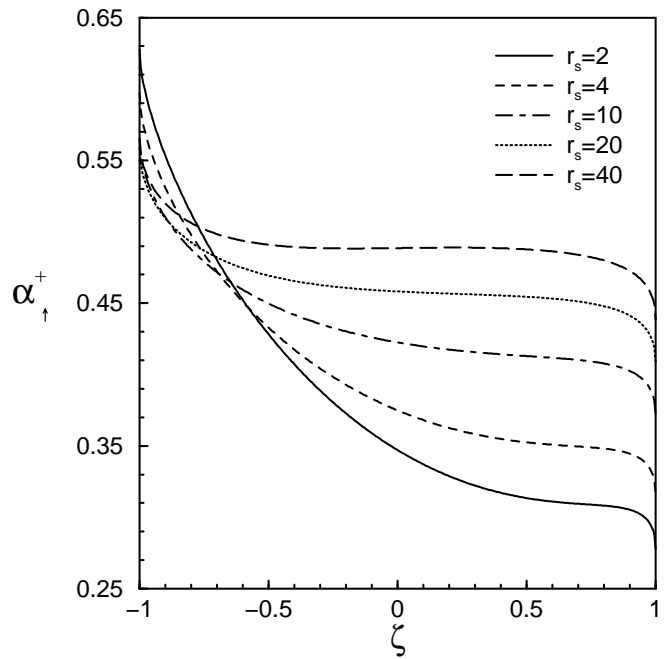}
\caption {Initial slope of the local field correction function
$\alpha^{+}_{\uparrow}$ as a function of the polarization
($\zeta$) for different values of $r_s$ as indicated in the
legend.} \label{fig:alphaplus}
\end{minipage}
\end{figure}

\begin{figure}
\begin{minipage}{\linewidth}
\includegraphics[width=\textwidth]{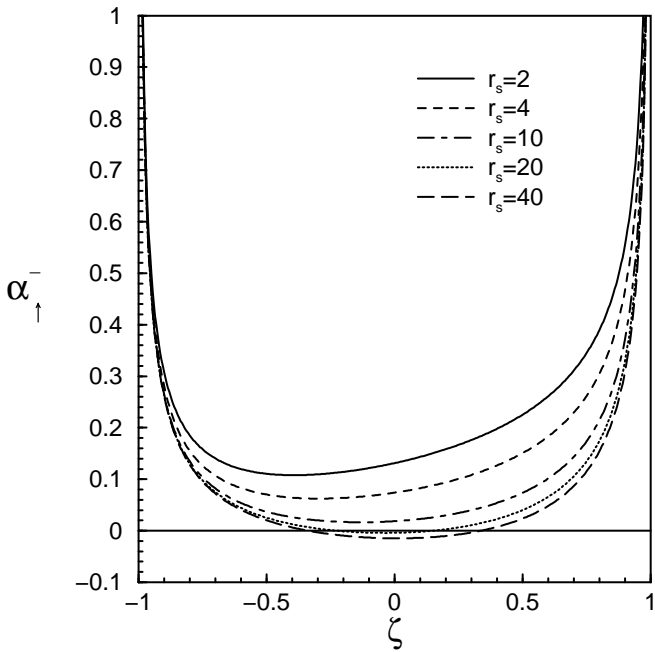}
\caption {Initial slope of the local field correction function
$\alpha^{-}_{\uparrow}$ as a function of the polarization
($\zeta$) for different values of $r_s$.} \label{fig:alphaminus}
\end{minipage}
\end{figure}

Any quantitative calculation of the effects of the
electron-electron interaction on many physical properties of
interest requires the knowledge of the correct form of the local
field correction functions. As we have already mentioned, 
obtaining the exact frequency and wave vector dependence of these
functions has been an elusive problem. Our first approximation is
to neglect the frequency dependence of the local field
corrections. Although the local field functions represent a
dynamical effect, they vary slowly on the scale of the Fermi
frequency, \cite{hol79,pede97} and it is acceptable to neglect
their frequency dependence.

The exact asymptotic values of $G^{\pm}_{\sigma}({\bf q},\omega)$
for large \cite{pol01} and small momentum\cite{mar02} at
zero frequency have been obtained previously and are summarized in
Table~\ref{tab:limitG}, where $\tilde{q}=q/k_F$ and $g(0)$ is the
two-particle correlation function at the origin. Note that the
values of $G^{x,\pm}_{\sigma}({\bf q}=0)$ included in the table
have been derived by considering only the contribution coming from
the electronic exchange energy. The proper expression for
$G^{\pm}_{\sigma}({\bf q}=0)$ incorporates also a combination of
derivatives of the correlation energy of the system,\cite{mar02} that
need to be added to the values given in Table~\ref{tab:limitG}.

If only the exchange contribution is considered, $G^+({\bf q}=
0)=G^-({\bf q}= 0)$ for the unpolarized electron gas. Using the
asymptotic values from Table~\ref{tab:limitG} and the previously
obtained expressions for the dielectric function
[Eq.~(\ref{eq:np.eps})] and magnetic susceptibility
[Eq.~(\ref{eq:np.sus})], we find that in the static limit both
response functions develop a pole for $r_s \ge \pi/\sqrt{2}\sim
2.221$. Note that a pole in the dielectric function does not
indicate an instability of the paramagnetic phase, it just implies
the appearance of a region in momentum space where the Coulomb
interaction is overscreened. \cite{Dolgov81} However, a pole in the
susceptibility points towards a spin density wave instability.
Since numerical calculations 
have shown the stability of the paramagnetic phase for $r_s
\lesssim 30$\cite{Tan89} it is crucial to include the correlation energy 
in the calculation of the local field factors. Including the correlations  
in $G^{\pm}_{\sigma}$ prevents the
occurrence of a spin/charge density wave instability in the
two-dimensional unpolarized electron gas.

Even though there are numerous calculations of the correlation
energy of an unpolarized electron gas, 
approximate expressions for the correlation energy of a polarized
electron gas are less numerous due to the fact that the magnetic
response functions are computationally harder to obtain.
\cite{MCsusc} For the two dimensional electron gas we use the
Monte Carlo calculation by Tanatar and Ceperley, \cite{Tan89}
where the correlation energy is expressed in the form of an
analytic interpolation formula: $E_c(r_s,\zeta)=E_c(r_s,\zeta=0) +
\zeta^2 (E_c(r_s,\zeta=1) -E_c(r_s,\zeta=0))$, where
$E_c(r_s,\zeta=0)$ and $E_c(r_s,\zeta=1)$ are approximated using a
Pad\'e scheme.

By including the contribution from the correlation energy, we
derived analytical expressions for  the initial slope of the local
field correction functions
$\alpha^{\pm}_{\sigma}=G^{\pm}_{\sigma}(\tilde{\bf q}\rightarrow
0)/(\tilde{\bf q})^{d-1}$, where $d$ is the dimension of the
system. These initial slopes are function of the density and the
polarization of the electron gas.  The behavior of $\alpha^{+}$
and $\alpha^{-}$ is quite different. Fig.~\ref{fig:alphaplus}
shows $\alpha^{+}_{\uparrow}$ for a two dimensional gas as
function of the polarization for several values of $r_s$. Note
that $\alpha^{+}_{\uparrow}$ is always positive, as it should be,
since local effects always decrease the bare electron charge
density. 
On the other hand, $\alpha^{-}_{\uparrow}$ depends strongly  on the
polarization for any electronic density as it can be seen in 
Fig.~\ref{fig:alphaminus} where $\alpha^{-}_{\uparrow}$ is displayed.
In our approximation $\alpha^{-}_{\uparrow}$
diverges for a fully polarized electron gas and, as a consequence,
$G^{-}_{\uparrow}({\bf q})$ becomes a constant and equal to its
value at ${\bf q}\rightarrow \infty$.
In addition, $\alpha^{-}_{\uparrow}$ becomes negative in diluted
systems and small polarizations. \cite{negalpha}
This behavior favors the stability of the paramagnetic phase.
Note that similar arguments apply to $\alpha^{\pm}_{\downarrow}$
due to the fact that 
$\alpha^{\pm}_{\downarrow}(\zeta)=\alpha^{\pm}_{\uparrow}(-\zeta)$.

\begin{table}[t]
\begin{minipage}{\linewidth}
\caption{Exact asymptotic values of the local-field factors
$G^{\pm}_{\sigma}({\bf  q},\omega=0)$ of a two dimensional
electron gas for large and small ${\bf q}$ at zero frequency. 
$G^{\pm}_{\sigma}({\bf q}\rightarrow
\infty)$ was calculated in Ref.~\onlinecite{pol01}.
The values of $G^{x,\pm}_{\sigma}({\bf q}=0)$ are
derived from Ref.~\onlinecite{mar02} considering only the
contribution coming from the electronic exchange energy.}
\begin{ruledtabular}
\begin{tabular} {||c|c||}
$G^{+}_{\uparrow}({\bf q}\rightarrow \infty)$
& $1-(1-\zeta)g(0)$\\
\hline $G^{-}_{\uparrow}({\bf q}\rightarrow \infty)$ &
$(1-\zeta)g(0)$\\
\hline $G^{x,+}_{\uparrow}({\bf q}\rightarrow 0)$ & $(\tilde{q}/(2
\pi)) \big\{ \zeta \sqrt{1+\zeta}+(2-\zeta) \sqrt{1-\zeta} \big\}$
\\
\hline 
$G^{-}_{\uparrow}({\bf q}\rightarrow 0)$ & $(\tilde{q}/(2
\pi)) \big\{ [(2
+\zeta)/\sqrt{1+\zeta}]+[\zeta/\sqrt{1-\zeta}]\big\}$ \\
\end{tabular}
\label{tab:limitG}
\end{ruledtabular}
\end{minipage}
\end{table}

\begin{figure}
\begin{minipage}{\linewidth}
\includegraphics[width=\textwidth]{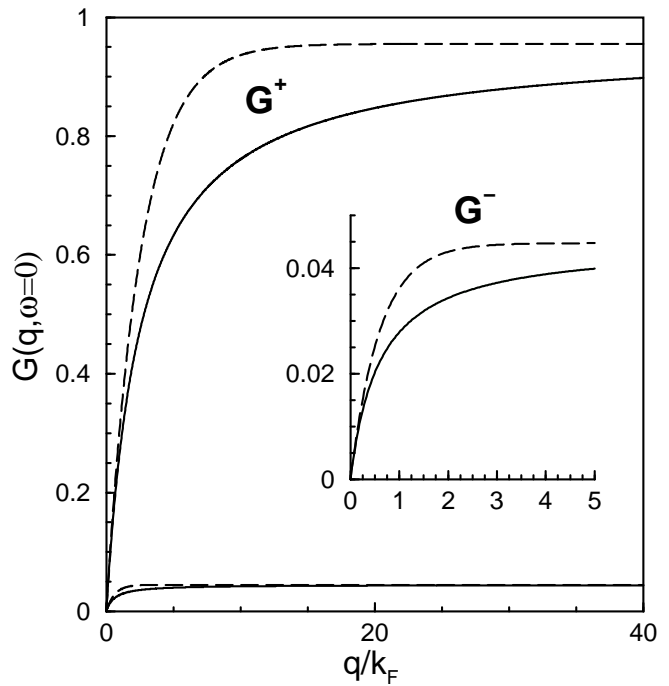}
\caption {Local field corrections $G^{\pm}_{\sigma}({\bf
q},\omega=0)$ versus normalized momentum ${\bf q}/k_F$ for a
two dimensional unpolarized electron gas with $r_s=4$. Results for
the rational fit (Eq.(~\ref{eq:posrat})) (solid curves) and for
the exponential fit (Eq. (~\ref{eq:posexp})) (dashed curves) are
displayed. The inset is a blow-up of the lower corner with the
$G^{-}$ functions.} \label{fig:2DGf0}
\end{minipage}
\end{figure}

\begin{figure}
\begin{minipage}{\linewidth}
\includegraphics[width=\textwidth]{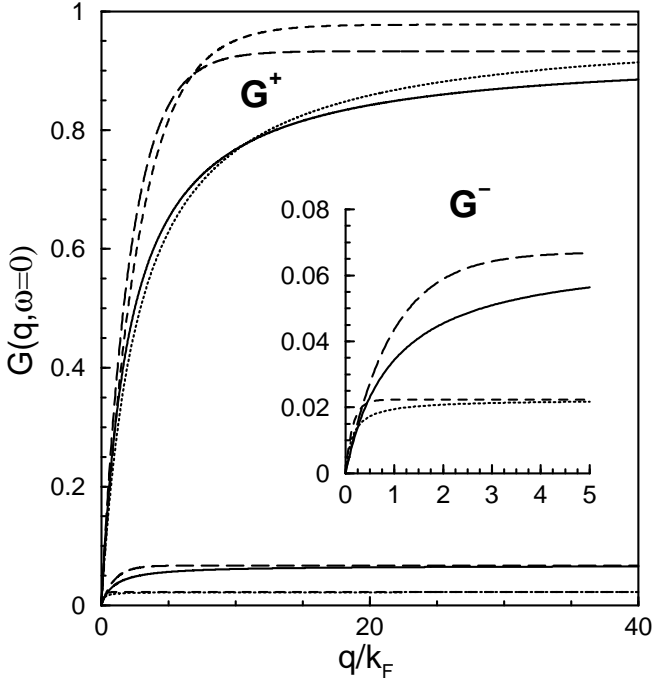}
\caption {Local field corrections $G^{\pm}_{\sigma}({\bf
q},\omega=0)$ versus normalized momentum ${\bf q}/k_F$ for a
two dimensional electron gas with $r_s=4$ and polarization
$\zeta=0.5$. Results for the rational fit
(Eq.(~\ref{eq:posrat})), $G^{\pm}_{\uparrow}$  (dotted curves) and
$G^{\pm}_{\downarrow}$ (solid curves) and for the exponential fit
(Eq. (~\ref{eq:posexp})), $G^{\pm}_{\uparrow}$  (dashed) and
$G^{\pm}_{\downarrow}$  (long-dashed) are displayed. The inset is
also a blow-up of the lower corner with the $G^{-}$ functions.}
\label{fig:2DGf5}
\end{minipage}
\end{figure}

General expressions  for $G^{\pm}_{\sigma}({\bf q})$ can be
interpolated between the large and small momentum limits. The
simplest way to interpolate $G^{+}({\bf q})$ for a regular metal
was first suggested by Hubbard: \cite{Hubb57}
\be
G^{+}({\bf
q})=\frac{\alpha {\bf q}^2}{1+\Big( \alpha/ G^{+}(\infty) \Big)
{\bf q}^2}
\ee

In a later work, Singwi and collaborators \cite{sing70,vash72}
remarked that their numerical results for $G^{+}({\bf q})$ in an
unpolarized three dimensional gas could be adequately fitted by
the simpler expression $G^{+}({\bf q})=A \Big( 1-e^{-B {\bf
q}^2}\Big)$, where the two fitting parameters are smoothly varying
functions of the electronic density. They noted that this
expression fits their results well at small and intermediate
values of momentum but not at larger values. However, $G^{+}({\bf
q})$ is relatively unimportant at large {\bf q}.
Similar conclusions have been reached for the unpolarized two
dimensional electron gas, \cite{bul96} where the argument of the
exponential is linear rather than quadratic. However, the
possibility that $G^{+}({\bf q})$ have a peak near $q=2 k_F$,
\cite{Wang84} as a residue of  the sharp peak in the exchange
potential, is  a long standing issue. Recent calculations show that
the inclusion of short-range correlations has the effect of
washing out this peak. \cite{pede97}

Approximate expressions for  $G^{-}({\bf q})$ are less numerous,
partly due to the fact that $G^{-}$ is related to the magnetic
response functions, which are computationally challenging.
\cite{MCsusc} In addition, it is known that for a three
dimensional gas, $G^{-}({\bf q})$ can not be a monotonic function
of momentum because its slope at ${\bf q}=0$ is positive but
asymptotically approaches a negative value at large momentum. Zhu
and Overhauser \cite{zhu86} suggested a simple rational function
with a maximum at ${\bf q}=2 k_F$.

Our approach shows that the initial slope of the local field
correction functions $G^{+}_{\sigma}({\bf q})$ is always positive.
In contrast, the initial slope of $G^{-}_{\sigma}({\bf q})$
could be negative for diluted systems and small values of the
polarization. In the large momentum limit, $G^{\pm}_{\sigma}({\bf
q}\rightarrow \infty)$ is always positive in two dimensions.
However, in three dimensions $G^{+}_{\uparrow}({\bf q}\rightarrow
\infty)$ might be negative for $\zeta \le -2/3$, and
$G^{-}_{\uparrow}({\bf q}\rightarrow \infty)$ is negative for all
values of the polarization if the electron density verifies $r_s
\ge 1.18$. \cite{largeq}

Given the diversity of behaviors in the limiting values of the
local field correction functions, we consider a general
interpolating scheme for all the cases. For a value of the
polarization $\zeta$ and of the electron density $r_s$, we
calculate the product $\alpha^{\pm}_{\sigma}
G^{\pm}_{\sigma}(\tilde{\bf q}\rightarrow \infty)$ where
$\alpha^{\pm}_{\sigma}$ has been defined previously. When
$\alpha^{\pm}_{\sigma} G^{\pm}_{\sigma}(\tilde{\bf q}\rightarrow
\infty)>0$, we use the following fitting expressions for the local
field factors: \ba \label{eq:posrat} G^{\pm}_{\sigma}(\tilde{\bf
q})= \frac {\alpha^{\pm}_{\sigma} \tilde{\bf q}^{d-1}} {1
+\Big(\alpha^{\pm}_{\sigma}/ G^{\pm}_{\sigma} (\infty) \Big)
\tilde{\bf q}^{d-1}}, \hspace{0.5cm} {\rm rational \hspace{0.2cm} fit}; \\
G^{\pm}_{\sigma}(\tilde{\bf q})= G^{\pm}_{\sigma} (\infty) \Big(
1- \exp^{-(\alpha^{\pm}_{\sigma}/G^{\pm}_{\sigma} (\infty) )
q^{d-1}}
\Big),\nonumber \\
{\rm exponential \hspace{0.2cm}  fit}.
\label{eq:posexp}
\ea

In the opposite case, $\alpha^{\pm}_{\sigma}
G^{\pm}_{\sigma}(\tilde{\bf q}\rightarrow \infty)<0$ we fit the
local field factor to  the simplest rational or exponential
function with a maximum at ${\bf q}=2 k_F$:

\ba
G^{\pm}_{\sigma}(\tilde{\bf q})= \frac {\alpha^{\pm}_{\sigma}
\tilde{\bf q}^{d-1} + \gamma^{\pm}_{\sigma} \tilde{\bf q}^{d+1} }
{1 +\Big(\gamma^{\pm}_{\sigma}/ G^{\pm}_{\sigma} (\infty) \Big)
\tilde{\bf q}^{d+1}}, \hspace{0.5cm} {\rm rational \hspace{0.2cm}fit}; \\
G^{\pm}_{\sigma}(\tilde{\bf q})= G^{\pm}_{\sigma} (\infty) \Big(
1- \exp^{\gamma^{\pm}_{\sigma} q^d-
(\alpha^{\pm}_{\sigma}/G^{\pm}_{\sigma} (\infty) ) q^{d-1}}
\Big), \nonumber \\
{\rm exponential\hspace{0.2cm} fit}.
\ea
where $\displaystyle
\gamma^{\pm}_{\sigma}=\frac{ \alpha^{\pm}_{\sigma} (d-1)/4} {2^d
(\alpha^{\pm}_{\sigma}/ G^{\pm}_{\sigma} (\infty))- (d+1)}$ for
the  rational fit, and $\displaystyle \gamma^{\pm}_{\sigma}=
\frac{d-1}{2 d}\frac{\alpha^{\pm}_{\sigma}} {G^{\pm}_{\sigma}
(\infty)}$ for the exponential fit.

Fig.~\ref{fig:2DGf0} shows the momentum dependence of the local
field correction functions, $G^{\pm}_{\sigma}({\bf q},\omega=0)$,
for an unpolarized two dimensional electron gas. We have chosen an
intermediate value for the electron density, $r_s=4$, which
corresponds to a density of 1.86 $\cdot 10^{10}$ cm$^{-2}$ in GaAs. 
It is clear that both
rational and exponential fittings show the same qualitative
behavior, although the exponential fitting function approaches the
large momentum limit faster than the rational one. Fig.
~\ref{fig:2DGf5} displays $G^{\pm}_{\sigma}({\bf q},\omega=0)$
versus normalized momentum for a polarization of $\zeta=0.5$ and
the same electronic density. It is clear how the local field
factors split in a polarized gas, although their functional
behavior is the same as that for zero polarization.

Finally, let us mention that the needed expression of the
two dimensional two-particle correlation function at the origin,
which is derived in the appendix, is:
\be
g(0)=\frac{1}{2(1+0.5857 r_s)^2}.
\ee

\section{\bf Conclusions}

Motivated by recent experimental developments in the physics of diluted
magnetic semiconductors, we devised a framework that explicitly
incorporates the spin degree of freedom of the itinerant carriers.
In our model, the static spins are integrated out
under the enlarged gyromagnetic ratio, and the itinerant
electrons are treated as a spin polarized gas
in the presence of a static magnetic field. 
Therefore, we focused on the itinerant carriers and the 
many body interactions
among them, which are modeled by using spin dependent local field
correction factors.

We found approximate expressions for the local field correction
functions in a two dimensional spin polarized gas by interpolating
their exact asymptotic values at small and large wavevectors. Our
results indicate that the overall behavior of the
local field corrections,
$G^{\pm}_{\s}({\bf q})$, does not depend on the exact interpolation
function used. Given
the density dependence of the local field functions, the values
derived in the paramagnetic electron gas approximation should
represent a realistic estimate irrespective of the source of the
polarizing magnetic field.

Using the expressions for the local field factors we calculated the
effective potentials between the two spin populations. In contrast
with the RPA approach, the inclusion of the local factors produces
a noticeable value of the interaction mediated by spin fluctuations. 
The interaction mediated by charge fluctuations is also greatly enhanced 
at small momenta. Charge and spin mediated interactions split when the 
electron gas is polarized, and, as a consequence, the  effective 
interaction between two electrons with spins $\sigma$ and $\sigma^{'}$ 
depends on the value of $\sigma$ and $\sigma^{'}$.

We also calculated the effective screened interaction, 
$\displaystyle V_{eff}= \frac{v({\bf q})}{\epsilon({\bf q},\omega=0)}$,
for the two approximate expressions of the local field correction
factors we used. We  found that both approximations produce
qualitatively equivalent potentials, which become overscreened for 
momenta less than a critical value, in agreement with other analytical
and numerical calculations.

In conclusion,
we believe that our approach  provides a realistic qualitative 
description of the 
paramagnetic phase of the diluted magnetic semiconductors.
Caution, however, should be
exercised in applying our calculation in the limit of $\zeta$
approaching unity, where the paramagnetic model breaks down and our
approximation might lead to singularities.

 {\bf Acknowledgments}

We gratefully acknowledge the financial support provided by the
Department of Energy, grant no. DE-FG02-01ER45897.

\appendix
\section{\bf Correlation function at the origin}

A realistic estimate of the two-particle correlation function at
the origin in a two-dimensional electron gas can be obtained by
following an idea developed in Ref. \onlinecite{Over95}. Since the
screened Coulomb interaction between electrons has radial
symmetry, a pair of electrons with opposite spin forms a singlet
with spatial wave function that depends only on  the magnitude of
the relative distance between them, $\Psi({\vec \rho}_1,{\vec
\rho}_2)=\Psi(\rho=|{\vec \rho}_1-{\vec \rho}_2|)$. By using
cylindrical coordinates and dropping the dependence on the angle
and the perpendicular coordinate we find that $\Psi(\rho)$ verifies the
following Schr\"odinger equation:
\be
-\frac{\hbar^2}{2 m^*} \Big (
\frac{d^2\Psi }{d \rho^2}+\frac{1}{\rho} \frac{d \Psi}{d \rho}
\Big ) + V(\rho) \Psi(\rho)=E  \Psi(\rho)
\ee
where  $V(\rho)$ is
the effective potential, and $m^*=m/2$ is the reduced mass. It is also
convenient to define $k$ such that $E= \hbar^2 k^2 /(2 m^*)$.

We approximate the screened Coulomb potential by the potential of
an electron that is surrounded by 
a circle of radius $r_s$ uniformly filled with screening charge
density $n e =e/(\pi r_s^2)$. Outside the
circle the charge is zero and so the effective potential,
\ba
V(\rho)=\frac{e^2}{r_s}\Big[ \frac{r_s}{\rho} -\frac{4}{\pi}
E\Big( \frac{\rho}{r_s} \Big) + \frac{4}{\pi} -1 \Big ],
\hspace{0.1in} \rho \leqslant r_s \nonumber \\
V(\rho)=0, \hspace{0.1in} \rho \geqslant r_s
\ea
where $E(x)$ is the complete elliptic  integral of the second kind.
For convenience, we introduce dimensionless variables, $x=\rho /(r_s a_B)$
and $q= k r_s a_B$, where $a_B=\hbar^2/m e^2$ is the Bohr radius.
The  Schr\"odinger equation becomes:
\ba
\frac{d^2 \Psi}{d x^2}+\frac{1}{x}\frac{d \Psi}{d x} + q^2 \Psi(x)=0
\hspace{0.05in};\hspace{0.05in} x \geqslant 1 \nonumber 
\ea
\ba
\frac{d^2 \Psi}{d x^2}+\frac{1}{x}\frac{d \Psi}{d x} + \Big\{
q^2 -r_s \Big[ \frac{1}{x} -\frac{4}{\pi}
E(x) + \frac{4}{\pi} -1 \Big ]\Big\} \Psi(x)=0; \nonumber
\ea
\ba
\hspace{1in} \ x \leqslant 1
\label{eq:Sch}
\ea

The general solution for $x \geqslant 1$ is given by 
$\Psi(x)= A J_0 (q x) + B N_0 (q x)$, where $A$ and $B$ 
are constants, $J_0$
is the Bessel function of order zero and $N_0$  the corresponding
Neumann's function. 
To find the solution inside the circle we make a Taylor expansion 
of $\displaystyle \Psi(x)=\sum_{n=0}^{\infty} \alpha_n x^n$. 
Since we are interested in the ground state of
the system, its energy is small and so is $q$. 
Then, we drop the $q^2$ term from
the differential equation, and arrive to the following recurrent
relation between the $\alpha_n$ coefficients:
\ba
n^2 \alpha_n =
r_s \Big\{ \alpha_{n-1} + \Big( \frac{4}{\pi}-3 \Big) \alpha_{n-2}
\nonumber \\
+ 2 \sum_{m=1}^{\infty} \Big[ \frac{(2m-1)!!}{2^m m!} \Big]^2
\frac{\alpha_{n-2m -2}}{2m-1}\Big\}
\label{eq:rec}
\ea

As a consequence of this recurrent relation,
every $\alpha_n$ is proportional to $\alpha_0$ and a function of
$r_s$: $\alpha_n=\alpha_0 F_n(r_s)$. 

In order to solve Eq.~(\ref{eq:Sch}) we match $\Psi(x)$ and its
derivative at $x=1$:
\ba
\Psi(x=1)& =& \alpha_0 G(r_s)= A J_0(q) + B N_0(q) \nonumber \\
\Psi^{'}(x=1)&=& \alpha_0 \widetilde{F}(r_s)= A J_0^{'}(q) + B N_0^{'}(q)
\ea
where $\displaystyle G(r_s)=\sum_{n=0}^{\infty} F_n(r_s)$ and 
$\displaystyle \widetilde{F}(r_s)=\sum_{n=0}^{\infty} n F_n(r_s)$.
By making an expansion in $q$ and keeping the higher orders, we 
arrive to the following relation:
\ba
\alpha_0 G(r_s)= A  + \frac{2}{\pi} B [\ln(q/2)+C] +O(q^2 \ln q) \nonumber \\
\alpha_0 \widetilde{F}(r_s)= A (-q/2)+ \frac{2}{\pi} B \Big[\frac{1}{q} +
\frac{q}{2}\Big( \frac{1}{2} -C \Big) \nonumber \\
-\frac{q}{2}\ln \Big( \frac{q}{2}\Big) \Big] + O(q^2 \ln q)
\ea
where $C$ is the Euler constant. Therefore, in the limit of small momentum
\be
\alpha_0 \propto \frac{1}{G(r_s)}+
\frac{\widetilde{F}(r_s)}{G(r_s)^2} q \ln q
\ee
Using the recurrent relation~(\ref{eq:rec}), we can obtain $G(r_s)$:
\ba
G(r_s)= 1 + r_s \Big\{\frac{1}{4} +\frac{1}{\pi} +\nonumber \\ 
+\sum_{m=1}^{\infty} \Big[ \frac{(2m-1)!!}{2^m m!} \Big]^2
\frac{1}{(2m-1)(2m+2)^2}\Big\}+ \nonumber \\
 +O(r_s^2)  \sim  (1 + 0.5857 r_s)
\ea

And, since the pair-pair correlation  $g(\rho=0)$
is proportional to the square of the wave function $\Psi(\rho=0)$,
\be
g(\rho=0)=\frac{1}{2}\frac{1}{[1+0.5857 r_s]^2}
\label{eq:pair-corr}
\ee

This formula agrees quite well with previous  calculations, as it can
be seen in Table~\ref{tab:g0} where the values of the correlation function 
at the origin, obtained using Eq.~(\ref{eq:pair-corr}), are compared
to the numerical calculation of Nagano et al. \cite{Nagano84} 
and the most recent estimate by Polini et al. \cite{Polini01}

\begin{table}
\begin{minipage}{\linewidth}
\caption{Two-particle correlation function at
the origin for various values of $r_s$. First column displays the values
obtained using Eq~(\ref{eq:pair-corr}), second one the numerical 
calculation by Nagano et al. and last column  the interpolation results
of Polini et al.}
\begin{ruledtabular}
\begin{tabular} {||c|c|c|c||}
$r_s$& Eq.~(\ref{eq:pair-corr}) & Nagano et al. & Polini et al.\\
\hline
0&0.5 &0.50&0.5 \\
0.5& 0.299&0.31&0.293\\
1 & 0.199&0.21&0.204\\
2 & 0.106&0.13&0.123\\
5& 0.032 &-&0.050\\
\end{tabular}
\label{tab:g0}
\end{ruledtabular}
\end{minipage}
\end{table}



\begin{thebibliography}{10}
\bibitem{same-spin} The correlation between same
spin electrons is considerably smaller than all the other
interactions, since same spin electrons are kept apart in
agreement with the Pauli principle.
\bibitem{MCsusc} Although there have been attempts to extract the
susceptibility from Monte Carlo simulations, large error bars are
generally associated with the results.
\end{thebibliography}
\end{document}